\documentclass[english,aps,manuscript,prb, superscriptaddress, reprint, footinbib]{revtex4-1}
\pdfoutput=1
\usepackage{lmodern}
\usepackage[T1]{fontenc}
\usepackage[latin9]{inputenc}
\usepackage{color}
\usepackage{float}
\usepackage{amsbsy}
\usepackage{amstext}
\usepackage{graphicx}

\makeatletter
 
 \@ifundefined{textcolor}{}
 {%
   \definecolor{BLACK}{gray}{0}
   \definecolor{WHITE}{gray}{1}
   \definecolor{RED}{rgb}{1,0,0}
   \definecolor{GREEN}{rgb}{0,1,0}
   \definecolor{BLUE}{rgb}{0,0,1}
   \definecolor{CYAN}{cmyk}{1,0,0,0}
   \definecolor{MAGENTA}{cmyk}{0,1,0,0}
   \definecolor{YELLOW}{cmyk}{0,0,1,0}
 }

\usepackage[colorlinks=true]{hyperref}
\usepackage[varg]{txfonts}
\usepackage{textcomp}

\makeatother

\usepackage{babel}

\begin{document}

\title{Rewritable nanoscale oxide photodetector}

\author{Patrick Irvin}

\author{Yanjun Ma}

\author{Daniela F. Bogorin}

\author{Cheng Cen}

\address{Department of Physics and Astronomy, University of Pittsburgh, Pittsburgh,
Pennsylvania 15260}

\author{Chung Wung Bark}

\author{Chad M. Folkman}

\author{Chang-Beom Eom}

\address{Department of Materials Science and Engineering, University of Wisconsin-Madison,
Madison, Wisconsin 53706}

\author{Jeremy Levy}

\email{jlevy@pitt.edu}

\address{Department of Physics and Astronomy, University of Pittsburgh, Pittsburgh,
Pennsylvania 15260}

\date{\today}

\maketitle
\textbf{Nanophotonic devices seek to generate, guide, and/or detect
light using structures whose nanoscale dimensions are closely tied
to their functionality.\citep{sirbuly_semiconductor_2005,agarwal_semiconductor_2006}
Semiconducting nanowires, grown with tailored optoelectronic properties,
have been successfully placed into devices for a variety of applications.\citep{wang_highly_2001,tian_coaxial_2007,fan_large-scale_2008}
However, the integration of photonic nanostructures with electronic
circuitry has always been one of the most challenging aspects of device
development. Here we report the development of rewritable nanoscale
photodetectors created at the interface between LaAlO$_{3}$ and SrTiO$_{3}$.
Nanowire junctions with characteristic dimensions 2-3 nm are created
using a reversible AFM writing technique.\citep{cen_nanoscale_2008,cen_oxide_2009}
These nanoscale devices exhibit a remarkably high gain for their size,
in part because of the large electric fields produced in the gap region.
The photoconductive response is gate-tunable and spans the visible-to-near-infrared
regime. The ability to integrate rewritable nanoscale photodetectors
with nanowires and transistors in a single materials platform foreshadows
new families of integrated optoelectronic devices and applications.}

The discovery of a quasi two-dimensional electron gas (q-2DEG) at
the interface between insulating oxides\citep{ohtomo_high-mobility_2004}
has accelerated interest in oxide-based electronics.\citep{Mannhart2008}
The interface between LaAlO$_{3}$ and SrTiO$_{3}$ undergoes an abrupt
insulator-to-metal transition as a function of the number of LaAlO$_{3}$
layers;\citep{thiel_tunable_2006} for structures at or near the critical
thickness ($d_{c}=3$ unit cells), the conductance becomes highly
sensitive to applied electric fields.\citep{thiel_tunable_2006} By
applying the electric field locally using a conducting AFM probe,
one can control this metal-insulator transition with resolution approaching
one nanometer.\citep{cen_nanoscale_2008,cen_oxide_2009} Devices such
as rectifying junctions\citep{bogorin_diode_2009} and transistors\citep{cen_oxide_2009}
can be created, modified, and erased with extreme nanoscale precision.
Additionally, as LaAlO$_{3}$ and SrTiO$_{3}$ are both wide-bandgap
insulators they are essentially transparent at visible wavelengths,
making it an interesting material system on which to search for photonic
functionality.

\begin{figure}
\begin{centering}
\includegraphics[width=80mm]{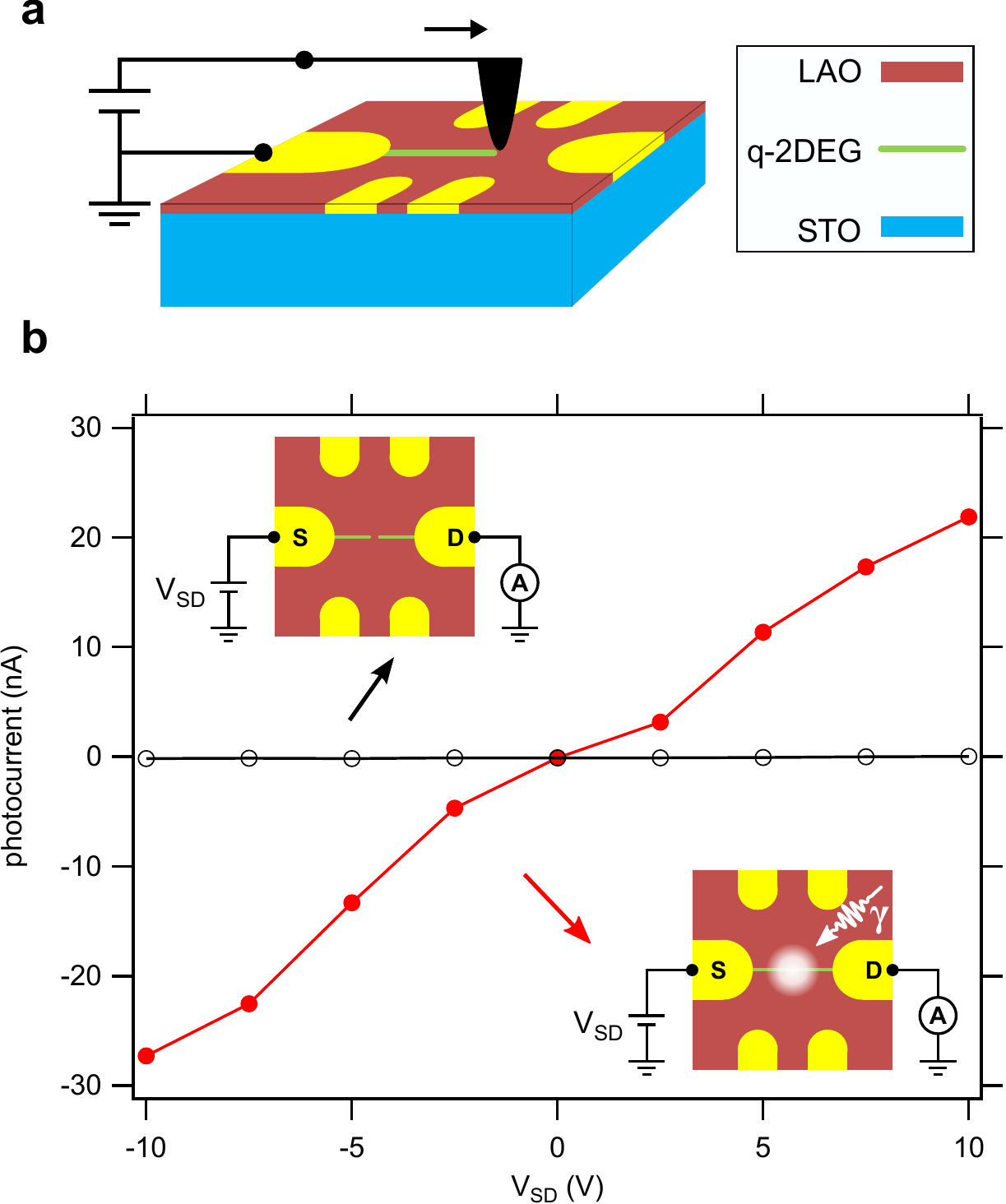}
\par\end{centering}

\caption{\textbf{\label{fig:Diagram-of-sample-1}Diagram of sample and photoresponse.}
\textbf{a,} A schematic illustrating how a conducting atomic force
microscope (AFM) tip writes a nanowire. \textbf{b,} Photocurrent vs.
$V_{SD}$ with (closed symbols) and without (open symbols) laser illumination.
$I=0.5$ mW/cm$^{2}$ and $\lambda=$633 nm. $T=300$ K.}

\end{figure}

In this letter we demonstrate the creation of nanophotonic devices
at the LaAlO$_{3}$/SrTiO$_{3}$ interface using conducting-AFM lithography
(Figure \ref{fig:Diagram-of-sample-1}a). Oxide heterostructures,
consisting of 3 units cells of LaAlO$_{3}$ on TiO$_{2}$-terminated
SrTiO$_{3}$ (ref. \onlinecite{Kawasaki1994}), are grown by pulsed-laser
deposition (see the Supplementary Information for details about growth
conditions). Following contact to the interface with low-resistance
Au electrodes, nanostructures are created at the LaAlO$_{3}$/SrTiO$_{3}$
interface by applying positive voltages to a conducting AFM tip. Nanoscale
insulating gaps are formed by {}``cutting'' these nanowires with
a negatively-biased AFM tip that passes over the nanowire. Electronic
nanostructures can be created with a high degree of precision and,
furthermore, are relocatable and reconfigurable. 

The nanostructures are characterized by applying a voltage bias to
a {}``source'' electrode ($V_{SD}$) and then measuring the current
from a {}``drain'' electrode ($I_{D}$). Optical properties are
measured by illuminating the sample with laser light; when the light
overlaps with the device a sharp increase in the conductance is observed.
A small, persistent photoconductive effect is observed under continuous
illumination (Supplementary Figure \ref{fig:SuppIdvsTime}). These
background effects are removed by modulating the laser intensity at
frequency $f_{R}$ using an optical chopper. The resulting photocurrent,
$i_{\gamma}$, is detected with a lock-in amplifier at $f_{R}$. Figure
\ref{fig:Diagram-of-sample-1}b shows a typical photocurrent response
as a function of $V_{SD}$. The photoconductive properties of these
nanodevices are mapped spatially using scanning photocurrent microscopy
(SPCM).\citep{Balasubramanian2005} A microscope objective ($NA=0.13$
or 0.73) mounted to a closed-loop, three axis piezo scanner focuses
the light and raster-scans it relative to the sample surface. The
resulting photocurrent is measured as a function of laser position.
To maximize nanostructure lifetimes,\citep{cen_oxide_2009} measurements
are performed in a vacuum of $<1$ mbar. To reduce the signal from
thermally-activated carriers, some experiments are performed in a
continuous-flow cryostat at $T=80$ K.

\begin{figure*}
\begin{centering}
\includegraphics[width=160mm]{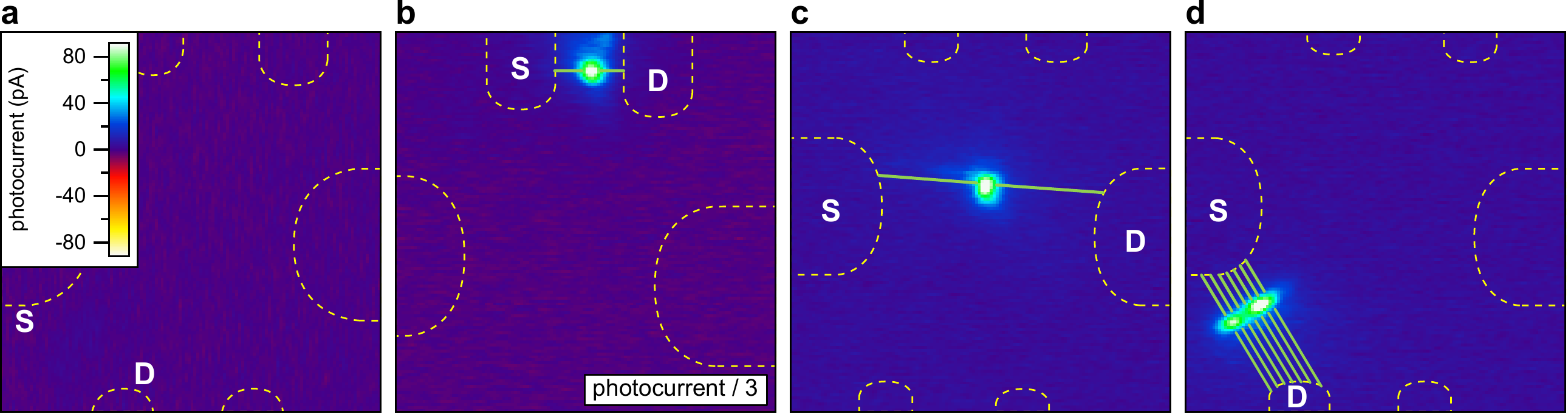}
\par\end{centering}

\caption{\label{fig:SPCM}\textbf{Scanning photocurrent microscopy (SPCM) images
of various nanostructures written at the LaAlO$\mathbf{\mathbf{_{3}}}$/SrTiO$_{\boldsymbol{3}}$
interface.} Images are $50\times50$ \textmu m$^{2}$. Dashed lines
indicate boundaries of areas where electrical contact is made to the
LaAlO$_{3}$/SrTiO$_{3}$ interface; solid lines indicate the locations
of nanowires. \textbf{a,} SPCM image of area before any nanostructures
are written. \textbf{b,} SPCM image for a nanowire junction written
close to a pair of electrodes. \textbf{c,} SPCM image formed after
erasing the previous nanowire and writing a second nanowire junction
from the electrodes. \textbf{d,} SPCM image for a set of seven parallel
wires with adjacent $w_{j}=2.5$ nm gaps; the separation between each
wire is 2 \textmu m. \textbf{a, b, $I\sim2.5$} kW/cm$^{2}$, $V_{SD}=0.5\textrm{ V}$.
\textbf{c, d, $I\sim3.8$ }kW/cm$^{2}$,\textbf{ }$V_{SD}=0.1\textrm{ V}$.
$T=300$ K.}

\end{figure*}

The photosensitivity of the devices written at the LaAlO$_{3}$/SrTiO$_{3}$
interface is spatially localized near the gap regions. An SPCM image
of the photocurrent between two electrodes that do not have a device
written between them shows a spatially diffuse photocurrent of less
than 2 pA (Figure \ref{fig:SPCM}a and Supplementary Figure \ref{fig:SuppNanophotonic-device-linecuts.}).\textbf{\emph{
}}The simplest nanophotonic device consists of a nanowire with a narrow
gap or junction. This device is created by first writing the wire
with an AFM tip bias of $V_{tip}=+10$ V, producing a nanowire with
width $w_{w}=2.5$ nm (Supplementary Figure \ref{fig:SuppWire width});
the junction is created by crossing the wire with $V_{tip}=-10$ V,
producing a gap with comparable width $w_{j}=2.5$ nm. The nanowire
junctions can be deterministically placed with nanometer-scale accuracy.
The SPCM image shows localized photocurrent ($\lambda=633$ nm and
$T=300$ K) detected in the region of the junction (Figure \ref{fig:SPCM}b).
The devices are erasable and reconfigurable: after measuring the device
shown in Figure \ref{fig:SPCM}b, the device was erased and a new
device was created farther from the electrodes (Figure \ref{fig:SPCM}c).
The photosensitivity of these devices can be optically modulated at
frequencies as high as 3.5 kHz and the response appears to be limited
by the RC time constant of the device (see the Supplementary Information).

More complex devices are readily created: a nanowire junction array
(Figure \ref{fig:SPCM}d) consisting of seven parallel wires spaced
2 \textmu m apart is drawn between the source and drain electrodes.
The nanowires are subsequently cut in a single stroke using a tip
bias of $V_{tip}=-10$ V, creating neighboring junctions of width
$w_{j}=2.5$ nm.\emph{ }The resulting photocurrent image shows the
expected stripe shape, demonstrating that the photocurrent signal
originates from all of the gaps. Even though the focused spot size
($\sim0.5$ \textmu m) is smaller than the line spacing, the individual
junctions are not separately distinguished. The photocurrent at the
interior junction is suppressed compared to the outermost junctions
due to the electrical screening by the nanowires. This result indicates
how carrier diffusion away from the junctions can also contribute
significantly to the overall photocurrent response. The extent of
the photoconductive spatial sensitivity appears to be of the order
of the spacing in this case ($\sim2$ \textmu m).

\begin{figure}
\centering{}\textsf{\includegraphics[width=85mm]{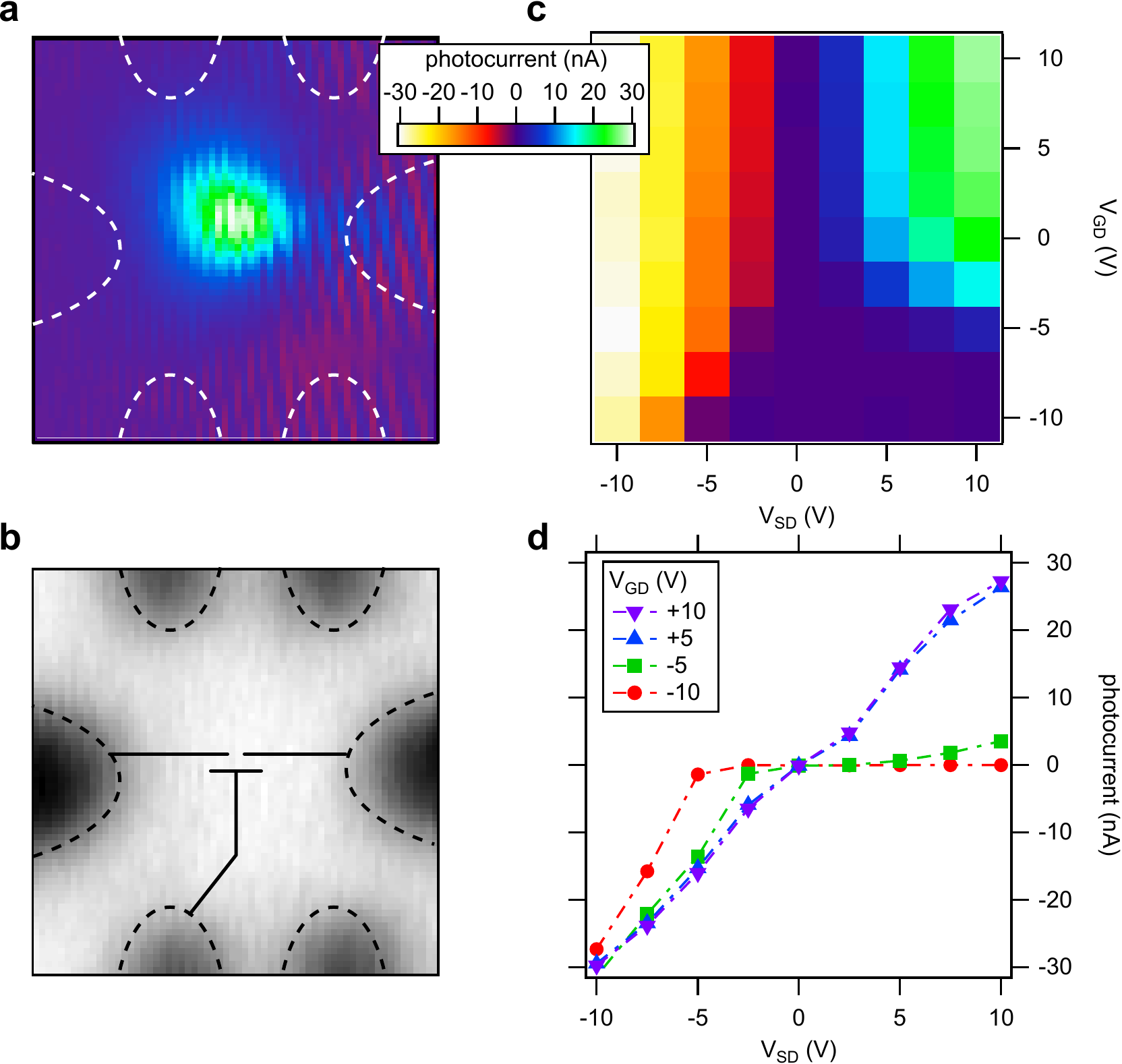}\caption{\label{fig:Tuning-633}\textbf{Three-terminal, nanoscale, locally
gateable photodetector.} \textbf{a,} SPCM image at $V_{SD},V_{GD}=+10$
V and \textbf{b,} simultaneously acquired reflectivity image. The
dashed lines show outline of electrodes and solid lines represent
nanostructures written with an AFM. Both the nanowire widths and gap
separations are exaggerated for clarity. Scan size is $50\times50$
\textmu m$^{2}$. \textbf{c,} Photocurrent as a function of $V_{SD}$
and $V_{GD}$. \textbf{d,} Photocurrent as a function of $V_{SD}$
plotted for different values of $V_{GD}$. $I\sim18$ W/cm$^{2}$.
$T=80$ K.}
}
\end{figure}

The functionality of these devices can be extended by adding an independent
gate electrode. Here, we adopt a geometry previously investigated
as a nanoscale transistor, i.e., a {}``SketchFET''\citep{cen_oxide_2009}.
The {}``gate'' electrode is written perpendicular to the existing
source-drain nanowire (Supplementary Figure \ref{fig:SuppNanophotonic-detector-geometry.}).
This bias $V_{GD}$ can be used to modify the source-drain conductance,
enabling conduction between source and drain for positive $V_{GD}$
and inhibiting it for negative $V_{GD}$. As for the two-terminal
wire with junction, photocurrent that is spatially localized near
the junction is observed where the device was written (Figure \ref{fig:Tuning-633}a).
A simultaneously acquired laser reflectivity image (Figure \ref{fig:Tuning-633}b)
does not show any observed signature of the nanophotonic detector,
such as changes in the absorption or scattering, which is also the
case for two-terminal devices. SPCM images are acquired for an array
of source and gate biases, $-5\textrm{\textrm{ V}}\leq V_{SD},V_{GD}\leq+5\textrm{\textrm{ V}}$
(Supplementary Figure \ref{fig:Supp3TerminalFitting}). The photocurrent,
measured as a function of $V_{SD}$ and $V_{GD}$ (Figure \ref{fig:Tuning-633}c.)
exhibits a polarity that is always the same sign as the $V_{SD}$,
irrespective of $V_{GD}$, indicating that there is negligible leakage
current from the gate to the drain. Furthermore, the photocurrent
is suppressed when both $V_{SD}$ is positive and $V_{GD}$ is negative,
demonstrating the ability of the gate electrode to tune the photoconductivity
in the source-drain channel.

\begin{figure}
\begin{centering}
\includegraphics[width=80mm]{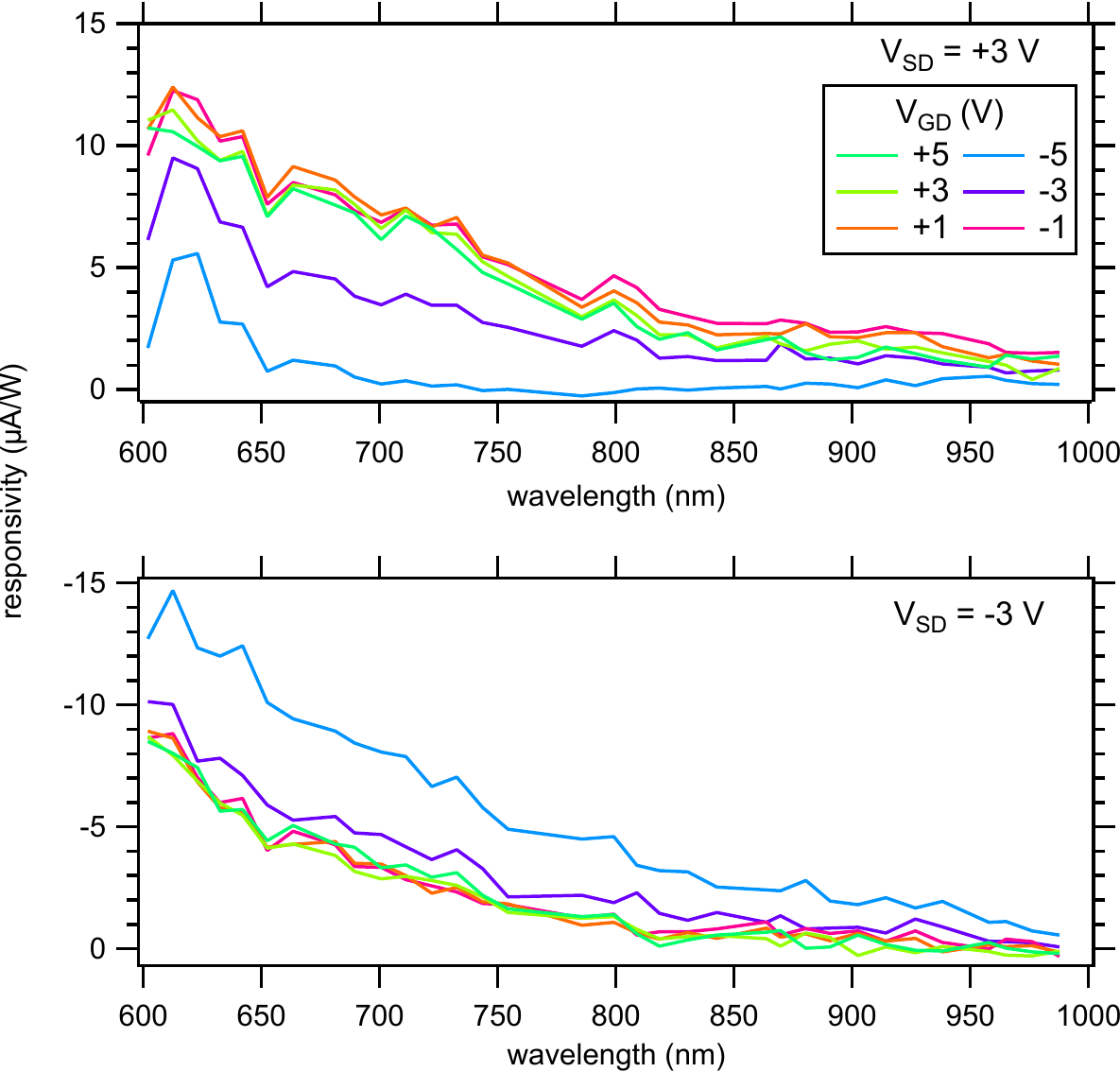}
\par\end{centering}

\caption{\label{fig:WhiteLightdependence.}\textbf{Gate-controlled spectral
response of photodetector.} Photocurrent as a function of $\lambda$
and $V_{GD}$ for $V_{SD}=+3$ V (top) and $V_{SD}=-3$ V (bottom).
$T=300$ K.}

\end{figure}

\begin{figure}
\begin{centering}
\includegraphics[width=80mm]{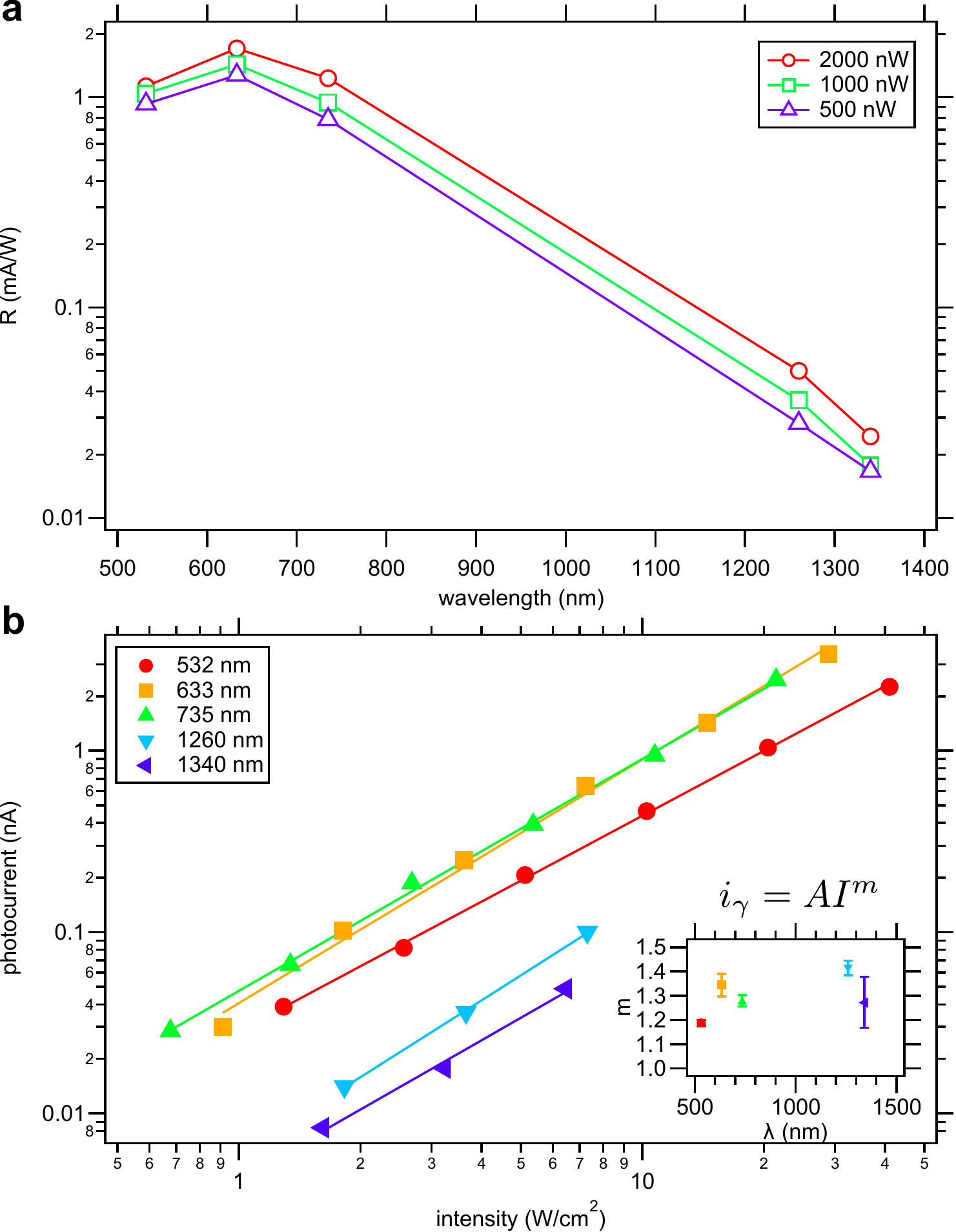}
\par\end{centering}

\caption{\label{fig:Wavelength-and-power}\textbf{Spectral sensitivity and
intensity dependence from visible to near-infrared wavelengths. a,}
Responsivity of photodetector from 532 nm to 1340 nm.\textbf{ b,}
Photocurrent versus optical intensity for different laser wavelengths.
Symbols are data and lines are fits. The inset shows the power-law
exponent $m$ as a function of wavelength. $V_{SD}=2$ V and $V_{GD}=0$
V. $T=80$ K. }

\end{figure}

To investigate the wavelength dependence of these devices, a pulsed,
mode-locked Ti:Sapphire laser is focused into a photonic crystal fiber
to provide tunable laser illumination over the continuous wavelength
range $600-1000$ nm (ref. \onlinecite{Dudley2006}). As the white
light source power varies with wavelength (Supplementary Figure \ref{fig:SuppSupercontinuum-power-vs.})
the normalized responsivity of the device ($i_{\gamma}/P$, where
$P$ is the laser power) is shown over this wavelength range (Figure
\ref{fig:WhiteLightdependence.}). Data points in the vicinity of
the pulsed laser source ($\lambda=780$ nm) are not shown because
of the high peak power and nonlinear effects in the sample (Figure
\ref{fig:Wavelength-and-power}). The spectral response is sensitive
to $V_{SD}$ and $V_{GD}$. At positive $V_{SD}$ the photodetector
response red-shifts as the gate bias is increased. The tuning of the
responsivity is enhanced for positive $V_{SD}$, which is consistent
with the behavior demonstrated in Figure \ref{fig:Tuning-633}. Measurements
taken as a function of temperature (Supplementary Figure \ref{fig:SuppSpectral-sensitivity-versus-T})
show a slight red-shift with increasing temperature.

In addition to the supercontinuum white light source, the optical
response is measured at a number of fixed wavelengths ranging from
the visible to near-infrared: 532 nm, 633 nm, 735 nm, 1260 nm, and
1340 nm. The response at these wavelengths is consistent with the
supercontinuum measurement in the range 600$-$1000 nm (Figure \ref{fig:Wavelength-and-power}a).
Remarkably, the photosensitivity extends to 1340 nm, the longest wavelength
investigated. A three-terminal device has similar tuning behavior
from $V_{SD}$ and $V_{GD}$ at 1340 nm and at visible wavelengths
(Supplementary Figure \ref{fig:SuppTuning-1340}). The intensity dependence
of the photocurrent exhibits power-law behavior (Figure \ref{fig:Wavelength-and-power}b),
$i_{\gamma}=AI^{m}$, where $i_{\gamma}$ is the photocurrent, $A$
is a proportionality constant, $I$ is the laser intensity, and $m\sim$
1.2$-$1.4. The super-linear scaling with laser intensity is similar
to other systems that are near a metal-insulator transition.\citep{Stockmann1969,Prezioso2009}

There are several possible mechanisms for the photoconductivity. The
direct and indirect bandgaps of SrTiO$_{3}$ (3.75 eV and 3.25 eV,
respectively\citep{van_benthem_bulk_2001}) are too large to explain
the visible to near-infrared photoresponse. Above-band photoexcitation
of SrTiO$_{3}$ can produce excitonic luminescence at visible wavelengths,
indicating the existence of mid-gap states.\citep{grabner_photoluminescence_1969,leonelli_time-resolved_1986,okamura_photogenerated_2006}
The most readily formed mid-gap states are associated with oxygen
vacancies, which are known to form during substrate preparation\citep{Kawasaki1994,Kareev2008,zhang:1466}
and growth of LaAlO$_{3}$/SrTiO$_{3}$ heterostructures.\citep{Herranz2007,kalabukhov_effect_2007,basletic_mappingspatial_2008,seo:082107}
Localized states just below the conduction band have been probed via
transport in SrTiO$_{3}$-based field-effect devices.\citep{shibuya_metal-insulator_2007}
Unintentional doping of SrTiO$_{3}$ substrates (e.g, Cr, Fe, or Al)
can also contribute states within the bandgap.\citep{feng_anomalous_1982}

Electrons occupying mid-gap states can be optically excited into the
conduction band using sub-bandgap light. The photoexcited electrons
are swept across the junction by the large electric field ($E=V_{SD}/w_{j}\sim100$
MV/cm), resulting in photocurrent. The spectral sensitivity we observe
(Figures \ref{fig:WhiteLightdependence.}, \ref{fig:Wavelength-and-power},
and Supplemental Figure \ref{fig:SuppSpectral-sensitivity-versus-T})
is consistent with previous optical measurements on oxygen vacancy-rich
samples.\citep{kalabukhov_effect_2007,Kareev2008,zhang:1466,seo:082107}
Along the nanowire and sufficiently far from the gap, photo-induced
current is negligible because the electric fields are screened or
otherwise sufficiently small.

The rewritable photodetectors presented here bring new functionality
to oxide nanoelectronics. For example, existing nanowire-based molecular
sensors\citep{cui_nanowire_2001} rely on the ability to bring the
analyte into contact with the sensing area of the detector. Here the
roles are reversed: a nanoscale photodetector can be placed in intimate
contact with an existing molecule or biological agent. The ability
to integrate optical and electrical components such as nanowires and
transistors may lead to devices that combine in a single platform
sub-wavelength optical detection with higher-level electronics-based
information processing.

\newpage
\begin{acknowledgments}
The authors gratefully acknowledge financial support from\emph{ }DARPA
(W911N3-09-10258) (J.L.) the Fine Foundation (J.L.), and the National
Science Foundation through grants DMR-0704022 (J.L.) and DMR-0906443
(C.B.E). 
\end{acknowledgments}
\bibliographystyle{apsrev4-1}
\bibliography{bib_OpticalSketchFET20100415}

\makeatletter
\setcounter{figure}{0}
\renewcommand{\thefigure}{\arabic{figure}}
\renewcommand{\figurename}{Supplementary Figure}
\makeatletter
\onecolumngrid
\newpage

\begin{center}
\textbf{\large Rewritable nanoscale oxide photodetector: Supplementary
Information}
\par\end{center}{\large \par}

\begin{center}
Patrick Irvin,$^{1}$ Yanjun Ma,$^{1}$ Daniela F. Bogorin,$^{1}$
Cheng Cen,$^{1}$ Chung\\
Wung Bark,$^{2}$ Chad M. Folkman,$^{2}$ Chang-Beom Eom,$^{2}$
and Jeremy Levy$^{1}$\\
$^{1}$\emph{Department of Physics and Astronomy,}\\
\emph{University of Pittsburgh, Pittsburgh, Pennsylvania 15260}\\
$^{2}$\emph{Department of Materials Science and Engineering,}\\
\emph{University of Wisconsin-Madison, Madison, Wisconsin 53706}
\par\end{center}

\pagenumbering{arabic}

\section*{Material Growth and Device Preparation}

Oxide heterostructures are grown by pulsed laser deposition with in
situ high pressure RHEED. Low miscut ($\sim0.05^{\circ}$) (001) SrTiO$_{3}$
substrates are treated by a modified BHF etch and annealed in oxygen
at 1000$^{\circ}$C for 2 hours to produce a TiO$_{2}$-terminated
and atomically smooth surface with single unit cell (uc) steps, as
verified by AFM inspection. Thin (3 uc) layers of LaAlO$_{3}$ are
deposited on SrTiO$_{3}$ at a temperature of $550-600^{\circ}$C
and oxygen pressure of $10^{-3}$ mbar. A laser with energy density
of 2 J/cm$^{2}$ and repetition rate of 3 Hz is used to ablate the
LaAlO$_{3}$ single crystal target.

In order to probe the interface of \textcolor{black}{LaAlO$_{3}$
and SrTiO$_{3}$, low-resistance electrodes are contacted directly
to the q-2DEG. Argon ion beam etching is used to mill $25$ nm deep
into the SrTiO$_{3}$. Electrodes are then formed by first sputter-depositing
a 2 nm Ti adhesion layer followed by $23$ nm of Au into the etched
region.}

Wires with width $w_{w}\sim2.5$ nm are written using an Asylum MFP-3D
AFM in contact mode. A positive bias voltage applied to the tip creates
conducting nanoregions directly below the AFM tip, while negative
voltages locally restore the insulating state. Nanostructures are
created under normal atmospheric conditions at room temperature; the
AFM is kept in a dark environment to suppress photoconduction in the
SrTiO$_{3}$ substrate ($E_{g}=3.2$ eV). The conductivity of the
q-2DEG is monitored with a Keithley picoammeter while writing: when
the wire is connected to both electrodes, a sharp increase in conductance
is observed. Similarly, the conductance is monitored while cutting
a wire. The distance over which the conductance drops is used to quantify
the wire width \citep{cen_nanoscale_2008}, as shown in Supplementary
Figure \ref{fig:SuppWire width}. Writing and erasing of nanowires
is reproducible for a given tip bias and oxide heterostructure.

\section*{Optical Modulation}

The photosensitivity of these devices can be optically modulated at
frequencies as high as 3.5 kHz, as shown in Supplementary Figure \ref{fig:SuppIdvsTime}.
The phase of the photocurrent is constant over the frequency range
probed while under illumination (ie, the {}``on'' state). In this
case the light lowers the resistance of the detector. The frequency
response of the device is limited by the RC time constant of the junction.
When the laser light is blocked ({}``off''), the detector is in
a high resistance state. In the off state we observe a phase shift
as the frequency is increased (Supplementary Figure \ref{fig:SuppIdvsTime}),
indicating we are indeed limited by the resistance.

\section*{Scanning Photocurrent Microscopy Image Analysis}

In order to extract a value for the photocurrent from a SPCM image,
the image is fit to a two-dimensional Gaussian of the form \[
f\left(x,y\right)=A_{0}+A_{1}\exp\left\{ \left(x-x_{0}\right)^{2}/2\sigma_{x}^{2}+\left(y-y_{0}\right)^{2}/2\sigma_{y}^{2}\right\} ,\]
where $A_{0}$ is the image offset, $A_{1}$ is the amplitude of the
2D peak, $x_{0}$ and $y_{0}$ are the peak offsets, and $\sigma_{x}$
and $\sigma_{y}$ are the peak widths. We then define the photocurrent
as $i_{\gamma}=A_{1}-A_{0}$.

\newpage

\begin{figure}[h]
\begin{centering}
\includegraphics[bb=0bp 0bp 372bp 207bp,clip,width=80mm]{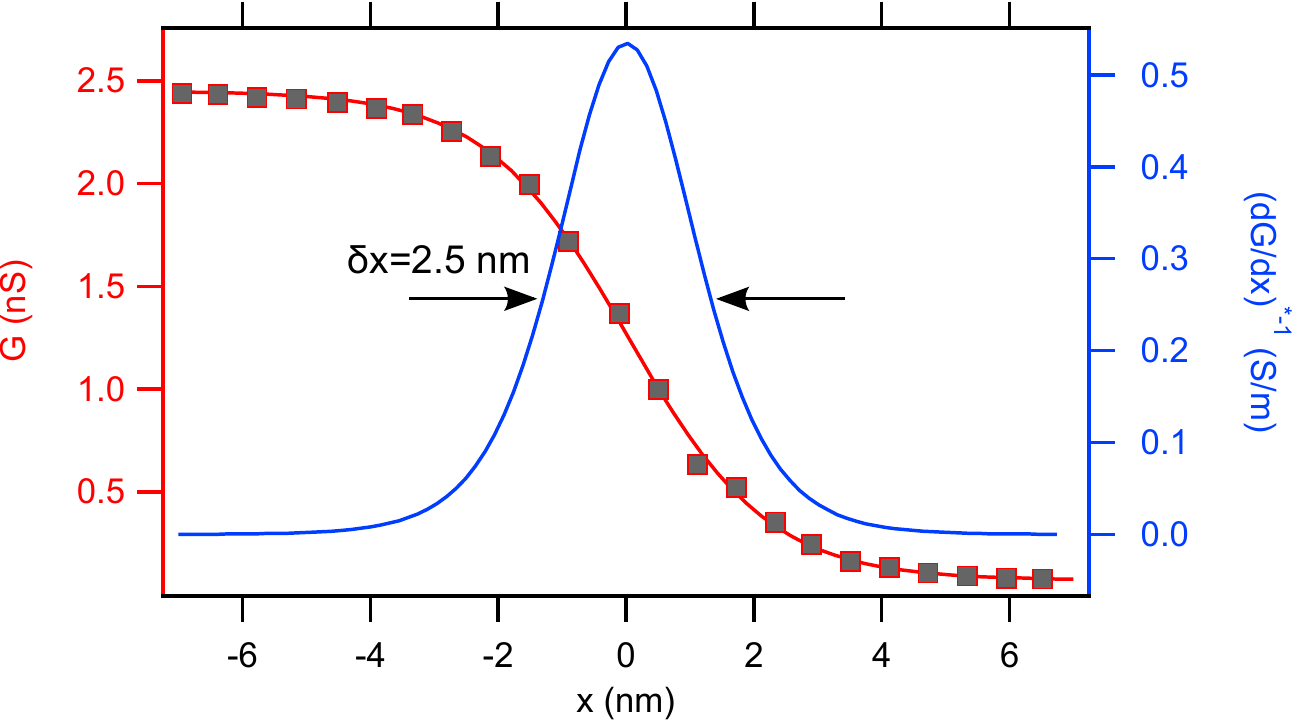}
\par\end{centering}

\caption{\label{fig:SuppWire width}\textbf{Width of wire.} Width of wire is
determined by moving a reverse-biased tip across the wire while monitoring
the conductance. The change in conductance is fitted to the function
$G(x)=G_{0}+G_{1}\tanh\left(x/h\right)$. Also plotted is the deconvolved
differential conductance $\left(dG/dx\right)^{*-1}$, from which we
determine the width of the nanowire.}

\end{figure}

\begin{figure}[h]
\begin{centering}
\includegraphics[bb=0bp 0bp 714bp 440bp,width=80mm]{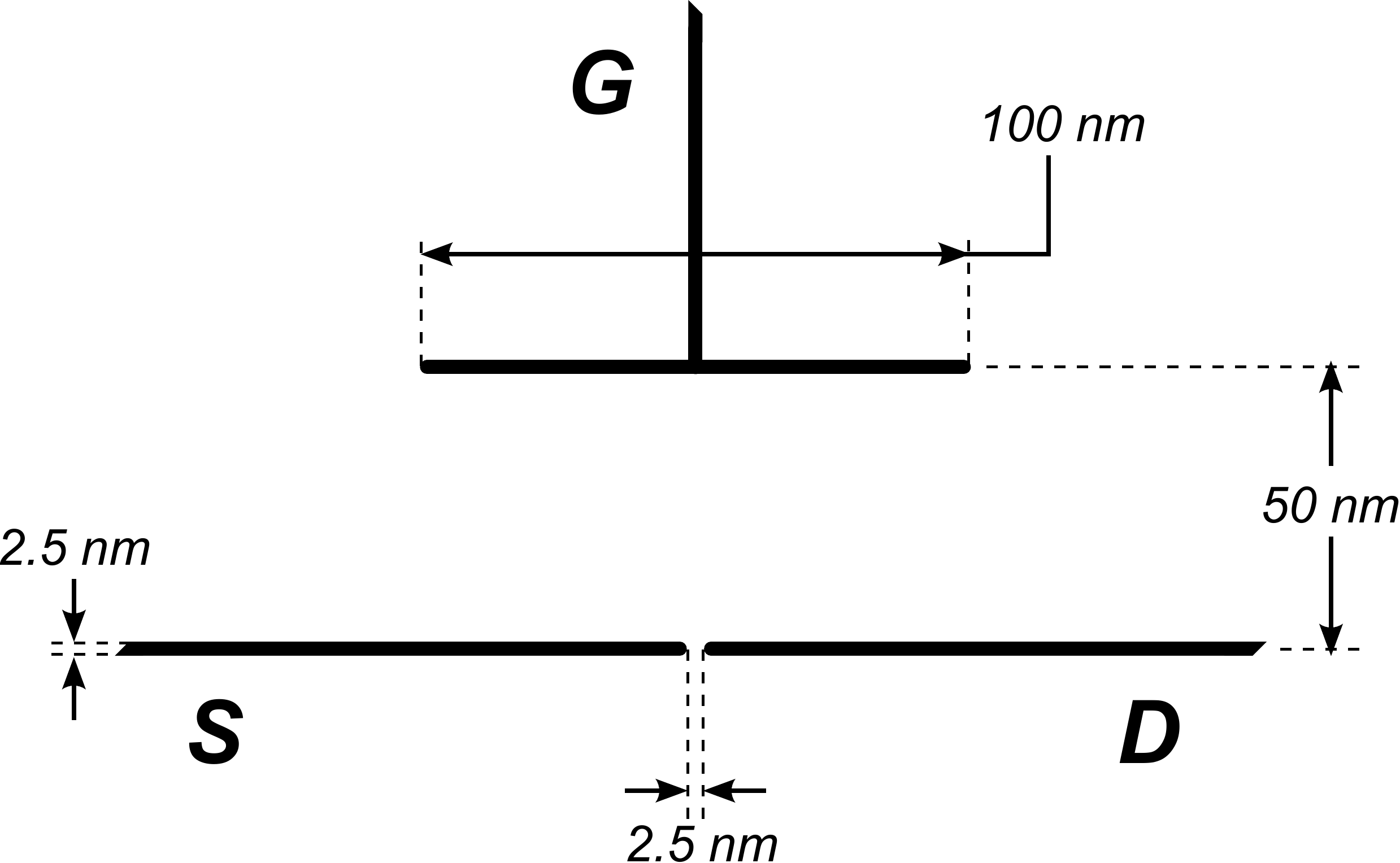}
\par\end{centering}

\caption{\label{fig:SuppNanophotonic-detector-geometry.}\textbf{Nanophotonic
detector geometry. }Typical geometry used for three-terminal devices.
The source-drain junction width $w_{j}$ is on the order of the size
of the wire width, $w_{w}=2.5$ nm. The gate electrode is positioned
50 nm from the source-drain wire. The {}``T'' shape helps ensure
a uniform electric field from the gate at the site of the junction.}

\end{figure}

\begin{figure}[h]
\begin{centering}
\includegraphics[width=80mm]{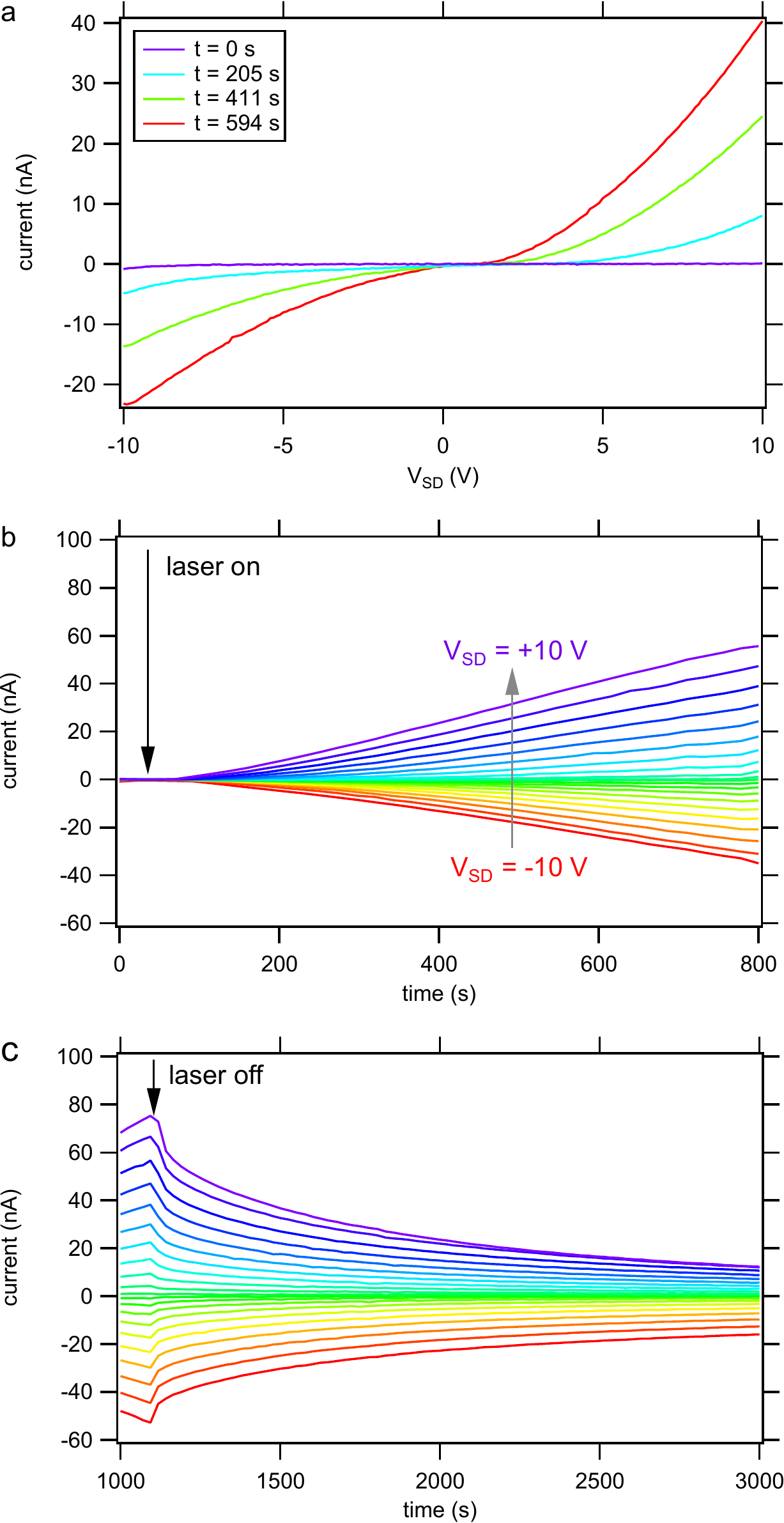}
\par\end{centering}

\caption{\label{fig:SuppDC-I vs. time}\textbf{DC current vs. time.} \textbf{a,}
DC current $I_{D}$ plotted as a function of $V_{SD}$. Each curve
is taken at a different time after turning on the laser illumination
($\lambda=633$ nm, $I=0.25$ mW/cm$^{2}$). \textbf{b, c,} DC current
as a function of time after laser is turned on (b) and off (c). Each
curve represents a different $V_{SD}.$ $T=300$ K.}

\end{figure}

\begin{figure}[H]
\begin{centering}
\includegraphics[width=80mm]{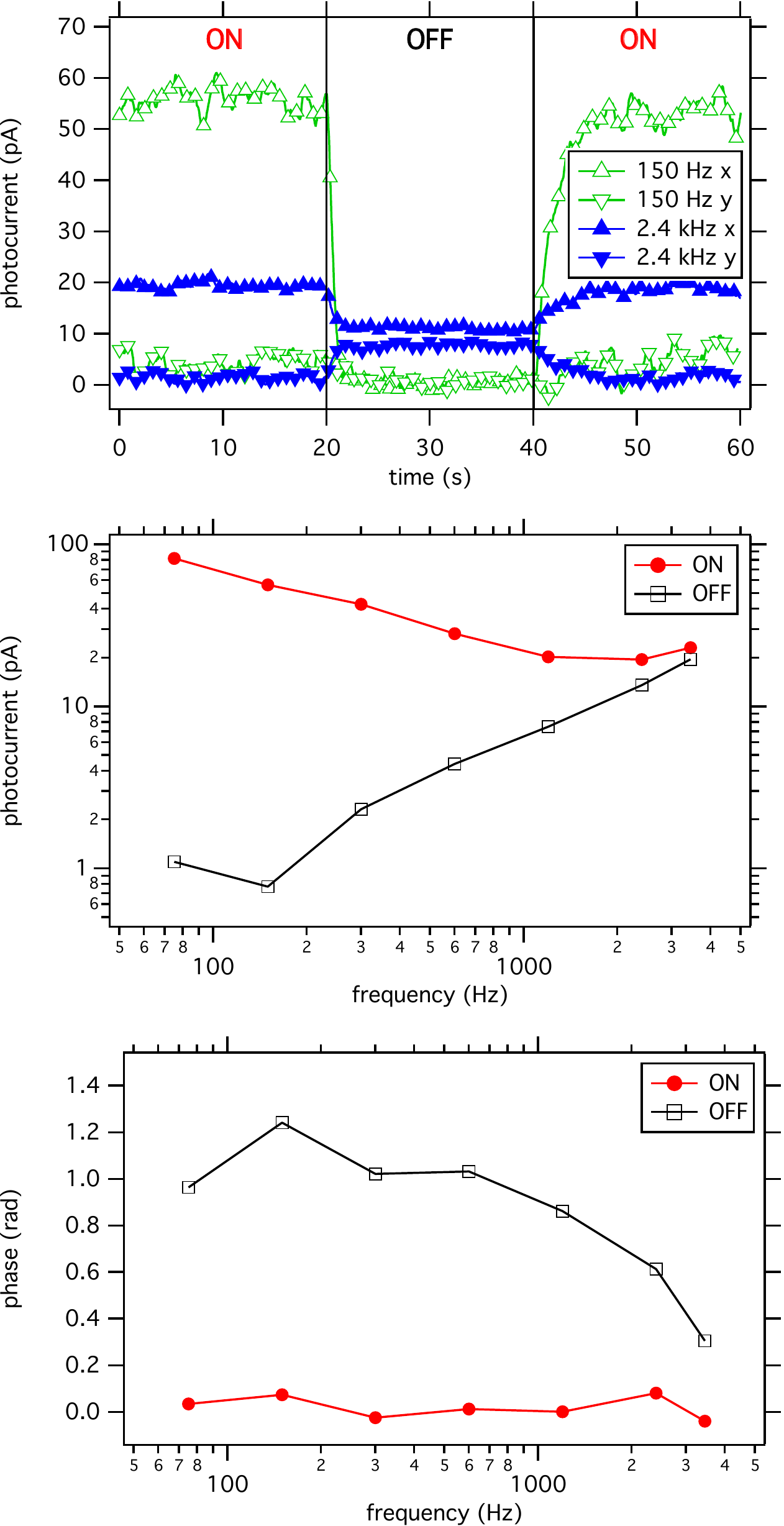}
\par\end{centering}

\caption{\label{fig:SuppIdvsTime}\textbf{Switching photosensitivity vs. frequency.
a, }Photocurrent versus time. Light that is initially on is blocked
at $t=20$ s and finally unblocked at $t=40$ s. The light is intensity
modulated by an optical chopper. Two modulation frequencies are shown:
$f_{R}=150$ Hz (open symbols) and $f_{R}=2.4$ kHz (closed symbols).
\textbf{b,} Photocurrent and \textbf{c,} Phase as a function of modulation
frequency in the on and off states. $I\sim0.5$ mW/cm$^{2}$. $T=80$
K.}

\end{figure}

\begin{figure}
\begin{centering}
\includegraphics[width=120mm]{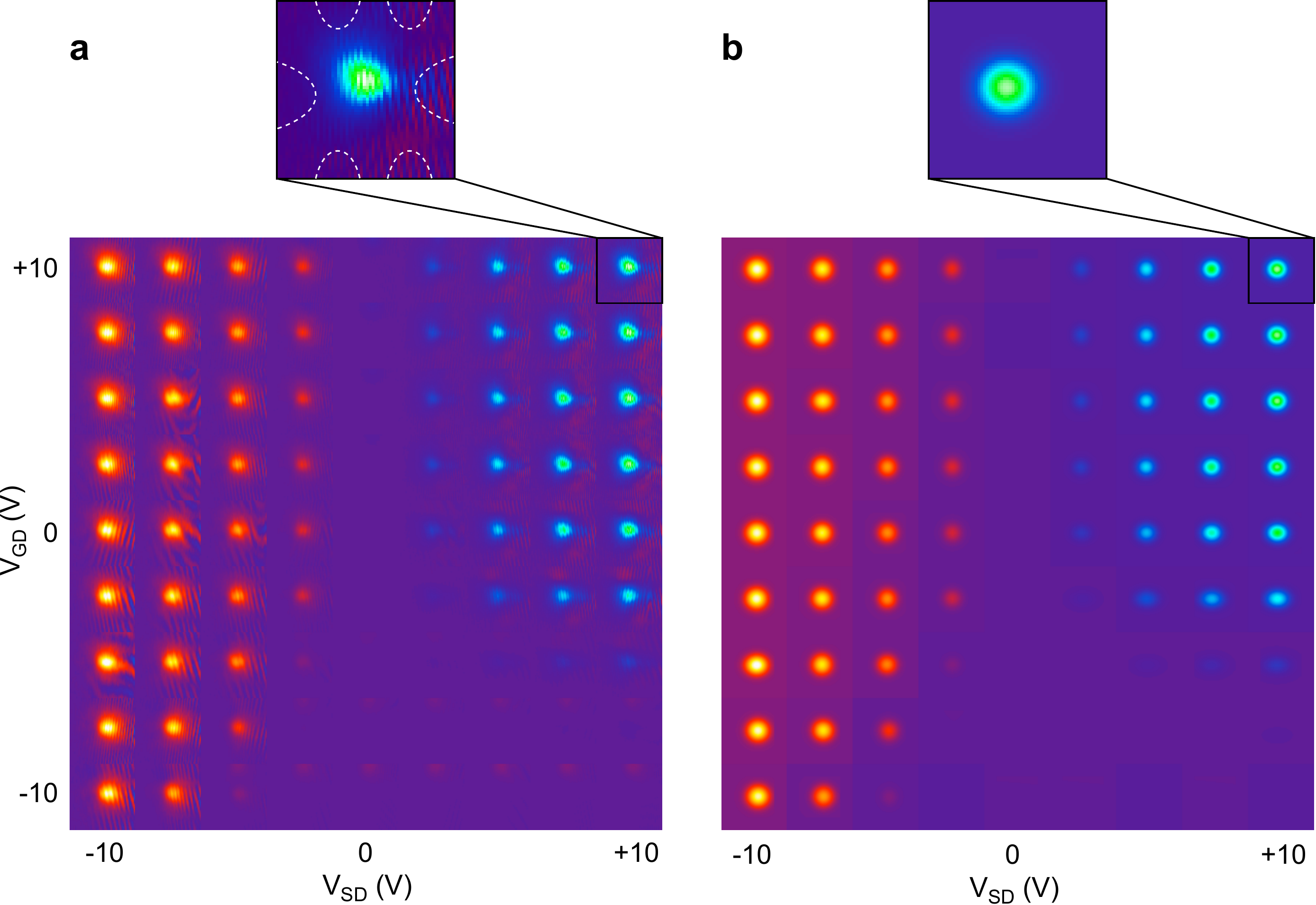}
\par\end{centering}

\caption{\label{fig:Supp3TerminalFitting}\textbf{Three terminal device analysis.}
\textbf{a,} Composite of SPCM images at different $V_{SD}$ and $V_{GD}$.
Inset shows SPCM image at $V_{GD}$, $V_{SD}=+10.5$ V. \textbf{b,}
A 2D Gaussian fit is performed on the data. The amplitude minus a
DC offset is plotted in Figures \ref{fig:Tuning-633}c and d.}

\end{figure}

\begin{figure}
\begin{centering}
\includegraphics[width=110mm]{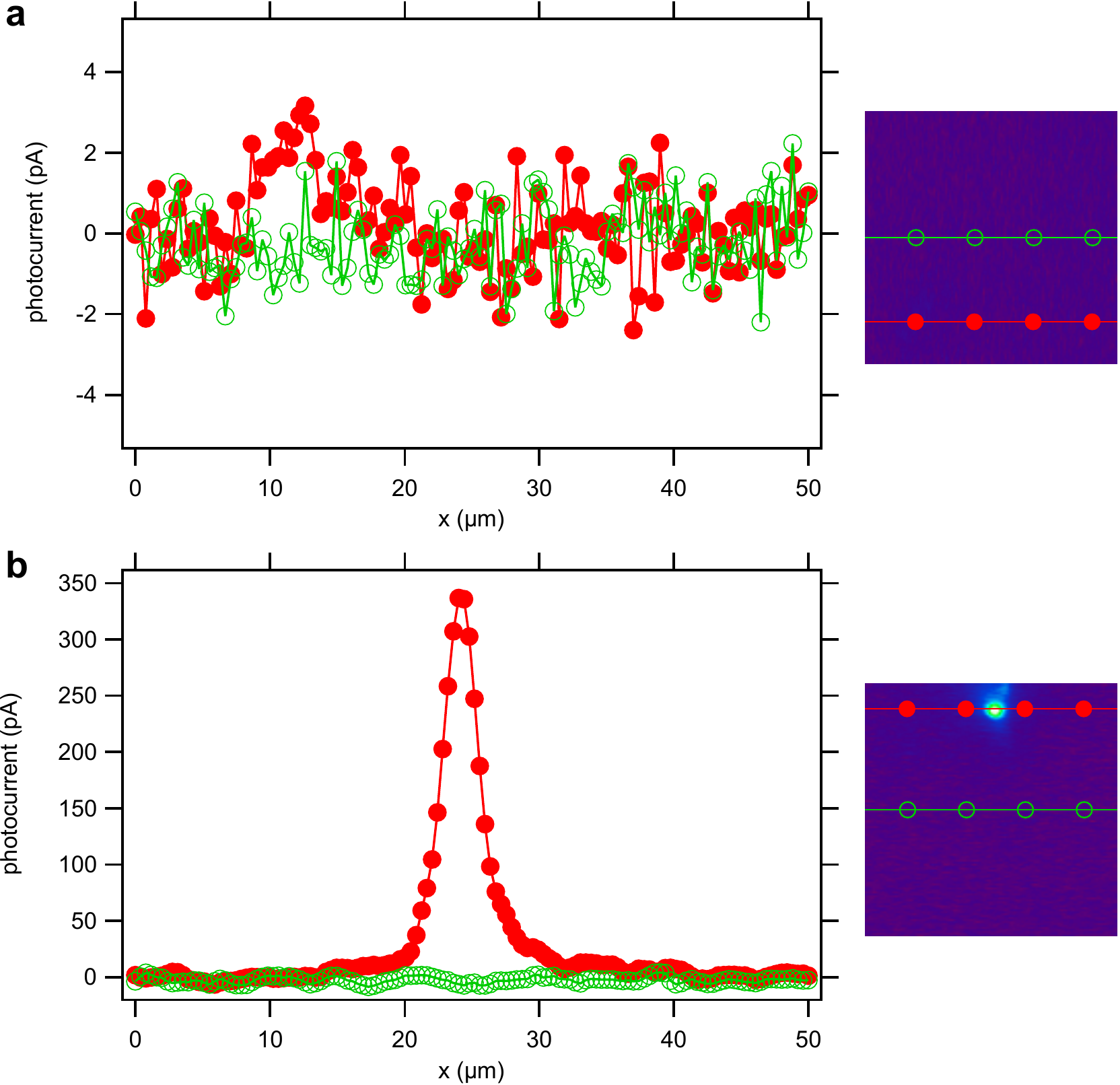}
\par\end{centering}

\caption{\label{fig:SuppNanophotonic-device-linecuts.}\textbf{Nanophotonic
device linecuts.} \textbf{a, b} Linecuts for the SPCM images shown
in Figures \ref{fig:SPCM}a and b, respectively. The green curves
(open symbols) show the signal away from the device while the red
curves (closed symbols) show linecuts through the device. For the
case where there is no device written (a), the red curve is drawn
between the two electrodes that were used to measure the photocurrent.}

\end{figure}

\begin{figure}
\begin{centering}
\includegraphics[width=80mm]{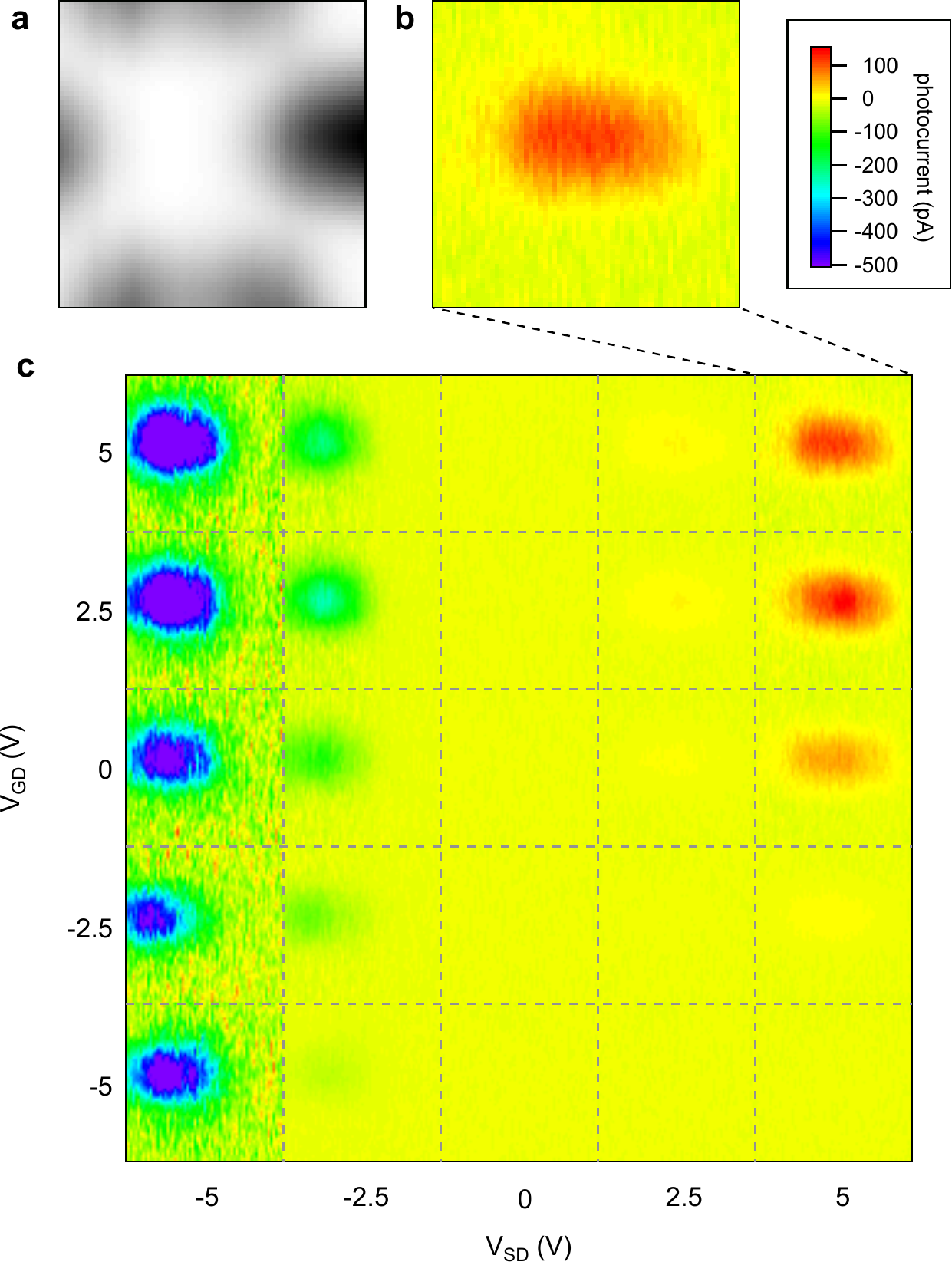}
\par\end{centering}

\caption{\label{fig:SuppTuning-1340}\textbf{Tuning the photoconductivity in
the near-infrared.} \textbf{a,} Reflectivity image and \textbf{b,}
simultaneously acquired SPCM image at $V_{SD}$,$V_{GD}=10$ V. \textbf{c,}
Composite image of SPCM images for an array of source and gate biases.
$\lambda=1340$ nm. $I\sim4$ W/cm$^{2}$. Data taken at $T=80$ K. }

\end{figure}

\begin{figure}
\begin{centering}
\includegraphics[width=80mm]{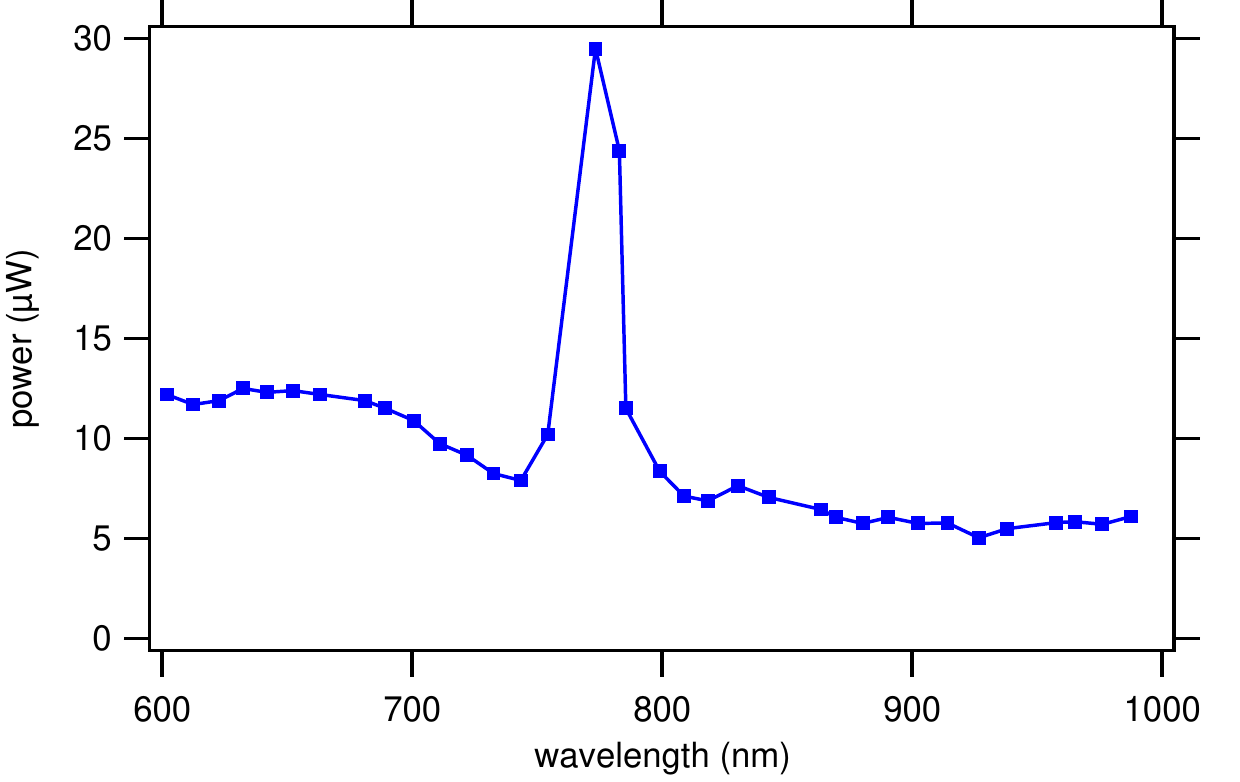}
\par\end{centering}

\caption{\label{fig:SuppSupercontinuum-power-vs.}\textbf{Supercontinuum laser
power vs. wavelength.} Supercontinuum light is generated by focusing
a pulsed, mode-locked Ti:Sapphire laser ($\lambda=780$ nm, $\delta t_{pulse}=150$
fs, $P=0.5$ W) into a photonic crystal optical fiber. After exiting
the fiber the light is filtered by a linear tunable bandpass filter
mounted on a motorized linear translation stage.}

\end{figure}

\begin{figure}
\begin{centering}
\includegraphics[width=160mm]{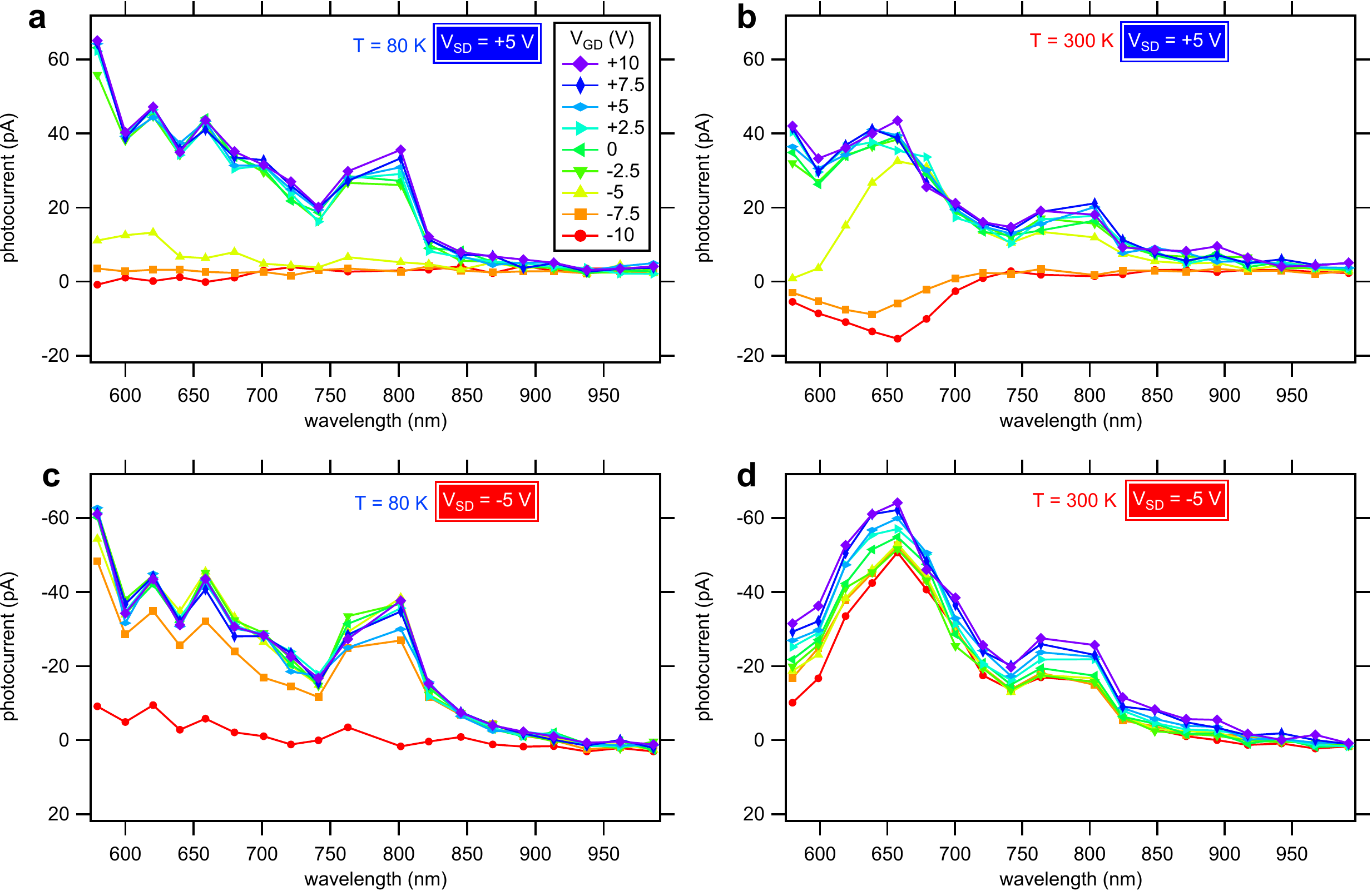}
\par\end{centering}

\caption{\label{fig:SuppSpectral-sensitivity-versus-T}\textbf{Temperature
dependence of spectral sensitivity. }Photocurrent plotted as a function
of wavelength at different bias conditions. \textbf{a, }$T=80$ K,
$V_{SD}=+5$ V. \textbf{b, }$T=300$ K, $V_{SD}=+5$ V. \textbf{c,
}$T=80$ K, $V_{SD}=-5$ V. \textbf{d, }$T=300$ K, $V_{SD}=-5$ V.
Laser power is shown in Supplementary Figure \ref{fig:SuppSupercontinuum-power-vs.}.}

\end{figure}

\end{document}